\documentstyle[prb,aps,multicol,epsf]{revtex}
\renewcommand{\narrowtext}{\begin{multicols}{2}
\global\columnwidth20.5pc}
\renewcommand{\widetext}{\end{multicols}
\global\columnwidth42.5pc} \multicolsep = 8pt plus 4pt minus 3pt

\begin{document}
\def \beq{\begin{equation}}
\def \eeq{\end{equation}}
\def \beqarr{\begin{eqnarray}}
\def \eeqarr{\end{eqnarray}}

\draft

\title{
Finite-temperature phase transitions in $\nu=2$ bilayer quantum Hall
systems }

\author {Min-Fong Yang${}^\dagger$ and Ming-Che Chang${}^*$
}

\address{
${}^\dagger$ Department of Physics, Tunghai University, Taichung,
Taiwan
\\${}^*$ Department of Physics, National Taiwan Normal
University, Taipei, Taiwan }

\date{\today}
\maketitle

\begin{abstract}
In this paper, the influence of an in-plane magnetic field
$B_\parallel$ on the finite-temperature phase transitions in
$\nu=2$ bilayer quantum Hall systems are examined. It is found
that there can exist two types of finite-temperature phase
transitions. The first is the Kosterlitz-Thouless (KT)
transitions, which can have an unusual {\em non-monotonic}
dependence on $B_{\parallel}$; the second type originates from the
crossing of energy levels and always increases with $B_\parallel$.
Based on these results, we point out that the threshold
temperature observed in the inelastic light scattering experiments
cannot be the KT transition temperature, because the latter shows
a totally different $B_{\parallel}$-dependence as compared with
the experimental observation. Instead, it should be the
level-crossing temperature, which we found agrees with the
$B_{\parallel}$-dependence observed. Moreover, combining the
knowledge of these two transition temperatures, a complete
finite-temperature phase diagram is presented.
\end{abstract}

\narrowtext

%\section{introduction}

Recent theoretical works predict that, besides the fully spin
polarized ferromagnetic phase (F) and the paramagnetic symmetric
or spin singlet (S) phase, a novel canted antiferromagnetic (C)
phase can exist in the filling factor $\nu=2$ bilayer quantum Hall
(QH)
systems.\cite{Zheng,DasSarma1,DasSarma2,Demler,KYang1,Brey,MacDonald}
In such a C phase, the electron spins in each layer are tilted
away from the external magnetic field direction due to the
competition between ferromagnetic ordering and singlet ordering.
Encouraging experimental evidence in support of the C phase has
recently emerged through inelastic light scattering
spectroscopy,\cite{Pellegrini1,Pellegrini2} transport
measurements,\cite{Sawada1,Sawada2} and capacitance
spectroscopy.\cite{Khrapai} In particular, it is observed for
certain samples in the inelastic light scattering
experiments\cite{Pellegrini1,Pellegrini2} that there is a
threshold temperature $T_{SDE}$ below which the spin conserved
spin-density excitation (SDE) mode ($\omega_0$ mode) seems to lose
all spectral weight. Because the $U(1)$ planar spin rotational
symmetry is spontaneously broken in the C phase, there should be a
finite-temperature Kosterlitz-Thouless (KT) transition with a
characteristic energy scale which is about the vortex-antivortex
binding energy.\cite{KT} It is claimed that the observed threshold
temperature is the predicted KT transition temperature $T_{KT}$ in
the C phase.\cite{DasSarma1,DasSarma2}

\begin{figure*}[h]
\centerline{\epsfxsize=7cm \epsfysize=7cm  \epsfbox{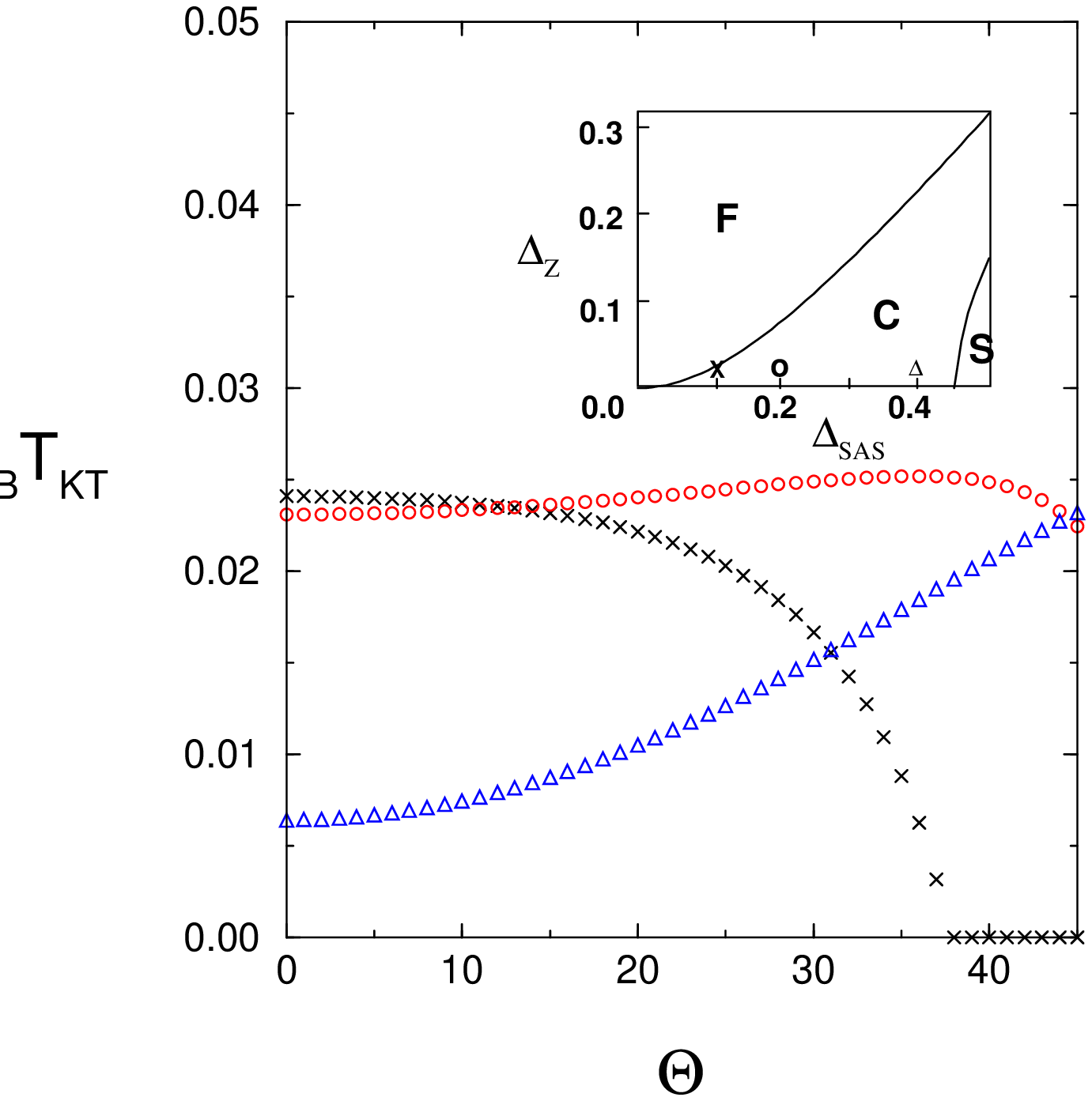} }
\caption{ $k_B T_{KT}$ as a function of the tilted angle $\Theta$ (in
unit of degree) of the applied magnetic field with a fixed $B_\perp$.
The energy unit is the intralayer Coulomb energy $e^2/\epsilon l$.
The Zeeman energy caused by $B_\perp$ is $\Delta_z=0.008$. The
interlayer separation is $d=1.45$ and the layer thickness is $b=1.0$.
Crosses, circles, and triangles correspond to $\Delta_{\rm SAS}=$
0.1, 0.2, and 0.4, respectively. Their locations in the
$\Delta_z-\Delta_{SAS}$ quantum phase diagram calculated by the
Hartree-Fock theory with $\Theta=0^\circ$ are shown in the inset.
Notice that the cross symbol represents the experimental sample of
Ref.[\protect\onlinecite{Pellegrini2}], and its location is very
close to the F-C phase boundary in the quantum phase diagram.
}\label{fig:KT}
\end{figure*}

\begin{figure*}[h]
\centerline{\epsfxsize=7cm \epsfysize=7cm \epsfbox{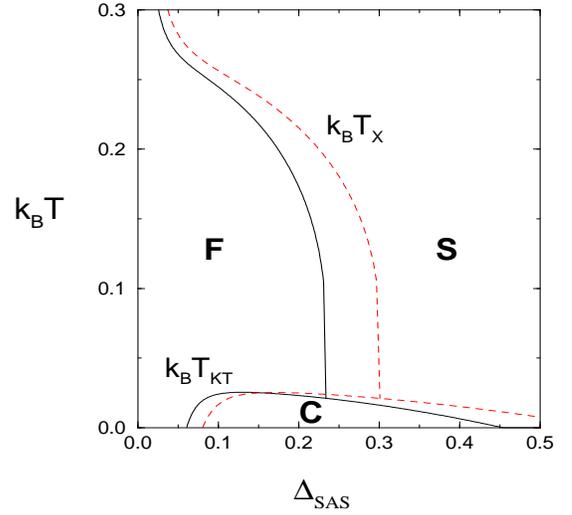} }
\caption{ $\nu$=2 bilayer phase diagrams at finite temperatures.
The energy unit is $e^2/\epsilon l$. The sample parameters,
$\Delta_z, d$, and $b$ are the same as in Fig.~\ref{fig:KT}. The
continuous lines are the phase boundaries for a perpendicular
magnetic field. The dotted lines are for a magnetic field tilted
by 30 degrees. } \label{fig:phase}
\end{figure*}

While the predicted value of the KT transition temperature
($T_{KT}\approx 1.8$ K in the Hartree-Fock
theory\cite{DasSarma1,DasSarma2}) is reasonably close to that of
the threshold temperature ($T_{SDE}\approx0.52$ K) in the
inelastic light scattering experiment under normal magnetic
fields,\cite{Pellegrini1} which seems to support the
identification between these two temperature scales, we point out
that such an interpretation meets trouble in the tilted magentic
field experiment\cite{Pellegrini2} for two reasons. First, it is
found in Ref.[\onlinecite{Pellegrini2}] that the threshold
temperature rises as the parallel magnetic field $B_{\parallel}$
increases. Nevertheless, we notice that (i) the sample used in the
experiment is located near the F-C phase boundary in the quantum
phase diagram (see the inset in Fig.~\ref{fig:KT}); and (ii) an
in-plane magnetic field {\it effectively} moves a sample even
closer to the F-C phase boundary.\cite{YC1} Hence the
symmetry-breaking order parameter and therefore $T_{KT}$ should be
reduced (c.f. Figs.~8 and 11 of Ref.[\onlinecite{DasSarma2}]),
rather than enhanced, when $B_{\parallel}$ increases. Thus the
physical content of these two characteristic temperatures should
not be the same. Second, it is questionable to regard the observed
disappearance of the $\omega_0$ mode at the threshold temperature
as the transition to the C phase, if one is reminded that the
spectral weight of the $\omega_0$ mode (which does not involve any
spin flip) is also greatly suppressed in the F phase, where almost
all spin-up (down) states are occupied (empty).\cite{note} When
temperature $T \gtrsim T_{KT}$ for the systems near the F-C
quantum phase boundary, it is expected that, although the
expectation value of the in-plane spin component vanishes and the
$U(1)$ planar spin rotational symmetry is restored, the spin
component $S_z$ along the direction of the external magnetic field
may still be nonzero. That is, these systems at $T \gtrsim T_{KT}$
should behave somewhat like the F phase at finite temperatures.
(See Fig.~\ref{fig:phase} for our finite-temperature phase
diagram.) Thus it needs higher temperatures for these systems to
loss all their spin polarizations, such that the $\omega_0$ mode
can be observed.

In this paper, the finite-temperature phase transitions in $\nu=2$
bilayer quantum Hall systems are investigated. As discussed above,
one had not yet reached the correct theoretical understanding for
the reported threshold temperature. Hence we focus our attention
on solving this issue. Since the aforementioned arguments are
quite general, the same qualitative results should be obtained
irrespective of which kind of approximation methods being
employed.\cite{YC2} For simplicity, we use the Hartree-Fock
approximation in the following. We show that the KT transition
temperature in the C phase\cite{DasSarma1,DasSarma2} can have an
unusual {\em non-monotonic} dependence on the tilted angle of the
applied magnetic field. That is, $T_{KT}$ can either rise or fall
as $B_{\parallel}$ is turned on, depending on whether the samples
are initially located near the C-S or the F-C phase boundary in
the quantum phase diagram. By using the sample parameters in the
tilted field experiment,\cite{Pellegrini2} we show that $T_{KT}$
{\it decreases} as the tilted angle increases. Thus the KT
transition scenario do fail to explain the
$B_{\parallel}$-dependence of the threshold temperature in
Ref.[\onlinecite{Pellegrini2}]. Instead, in order to link with the
observed threshold temperature, we propose another characteristic
temperature $T_X$ caused by the crossing of energy levels, since
its variation with respect to $B_{\parallel}$ agrees qualitatively
with the reported threshold temperature. Based on the dependence
of $T_{KT}$ and $T_X$ on the tunneling-induced
symmetric-antisymmetric energy gap, a complete finite-temperature
phase diagram is shown.

%\section{Kosterlitz-Thouless transition}

As shown in Ref.~[\onlinecite{DasSarma2}], the KT transition
temperature is estimated to be\cite{note1} $T_{KT} \approx 0.9
\rho_s/k_B$, with the spin stiffness $\rho_s =c_A\, \rho_s^A +
c_E\, \rho_s^E$, where
\begin{equation}\label{rho2}
\rho_s^{A/E} ={l^2\over4\pi}\sum_p p^2\, V_{A/E}(p,0),
\end{equation}
and the analytical forms of the constants $c_A$ and $c_E$ can be
written down explicitly by minimizing the Hartree-Fock variational
energy functional directly:\cite{MacDonald} \widetext
\begin{eqnarray}\label{extremum}
c_A&=& \frac{\Delta_{\rm SAS}^2 [ (\Delta_{\rm
SAS}^2-\Delta_z^2)^2-(2U_{-} \Delta_z)^2 ][(2U_{-} \Delta_{\rm
SAS})^2 - (\Delta_{\rm SAS}^2-\Delta_z^2)^2]}{(2U_{-})^2 (\Delta_{\rm
SAS}^2-\Delta_z^2)^4}, \\ c_E&=&\frac{[(2U_{-} \Delta_z)^2 -
(\Delta_{\rm SAS}^2-\Delta_z^2)^2][\Delta_z^2 (\Delta_{\rm
SAS}^2-\Delta_z^2)^2 - \Delta_{\rm SAS}^2 (2U_{-} \Delta_z)^2 ]}
{(2U_{-})^2 (\Delta_{\rm SAS}^2-\Delta_z^2)^4}.
\end{eqnarray}
\narrowtext Here $\Delta_{\rm SAS}$ is the tunneling-induced
symmetric-antisymmetric energy separation, $\Delta_z$ is the
Zeeman energy, and $U_{-} = (U_{A}-U_{E})/2$ with $U_{A/E}=\sum_p
V_{A/E}(p,0)$ being the exchange energy of the
intralayer/interlayer Coulomb interaction. The matrix elements
$V_{A/E}(p_1,p_2)$ of the intralayer/interlayer Coulomb
interaction are
\begin{equation}
V_{A/E}(p_1,p_2) ={1\over\Omega}\sum_{\bf q} v_{A/E}(q)
\delta_{p_1,q_y} e^{-q^2 l^2/2} e^{iq_X p_2 l^2},
\end{equation}
where $\Omega$ is the area of the system. $v_A(q)=(2\pi
e^2/\epsilon q)F_A(q,b)$ and $v_E(q)=v_A(q)F_E(q,b)e^{-qd}$ are
the Fourier transforms of the intralayer and the interlayer
Coulomb interaction potentials. $\epsilon$ is the dielectric
constant of the system, and $d$ is the interlayer separation. We
have also included the finite-well-thickness correction by
introducing the form factor $F_{A/E}(q,b)$ in the
intralayer/interlayer Coulomb potential, where
$F_A(q,b)=2/bq-2(1-e^{-qb})/b^2q^2$,
$F_E(q,b)=4\sinh^2(qb/2)/b^2q^2$, and $b$ is the width of a
quantum well.\cite{YC1,form} Since we know $\rho_s$ exactly within
the microscopic Hartree-Fock approximation, the KT transition
temperature can be easily determined. As shown in
Refs.[\onlinecite{DasSarma1,DasSarma2}], the KT transition
temperature along with the symmetry-breaking order parameter drops
continuously to zero as the phase boundaries are approached from
within the C phase.

Now we consider the tilted magnetic field case, where a parallel
magnetic field $B_\parallel$ and a perpendicular field $B_\perp$
both appear with the tilted angle
$\Theta=\tan^{-1}(B_\parallel/B_\perp)$. The effect of the
parallel magnetic field on $T_{KT}$ can be incorporated by the
following replacements:\cite{YC1,note2}
\begin{eqnarray}
\Delta_{\rm  SAS}&\rightarrow&{\bar \Delta}_{\rm  SAS}= \Delta_{\rm
SAS} e^{-Q^2 l^2 /4}, \nonumber\\
\Delta_z&\rightarrow&{\bar\Delta}_z=\Delta_z \sqrt{ 1 + (
B_\parallel/B_\perp )^2}, \label{new}\\
V_E(p_1,p_2)&\rightarrow&{\bar V}_E(p_1,p_2)=V_E(p_1,p_2)e^{\pm iQ
p_1 l^2}, \nonumber
\end{eqnarray}
with $Q=B_\parallel d/B_\perp l^2$ and the magnetic length
$l=\sqrt{\hbar c/eB_\perp}$. In Fig.~\ref{fig:KT} we show the
transition temperature $T_{KT}$ as a function of the tilted angle
$\Theta$ for some typical sample parameters. Since it is
relatively easy to tune $\Delta_{\rm SAS}$ in fabrication, we vary
its value with other system parameters being fixed. Three possible
situations are depicted in Fig.~\ref{fig:KT}: (i) if the system
begins in the C phase and near the C-S phase boundary (triangle),
then $T_{KT}$ (and $\rho_s$) grows as $B_\parallel$ is turned on;
(ii) if the system is in the C phase and near the F-C phase
boundary (cross), then $T_{KT}$ (and $\rho_s$) is reduced as
$B_\parallel$ is turned on; (iii) when the system lies between the
two phase boundaries (circle), $T_{KT}$ (and $\rho_s$) can have an
unusual {\em non-monotonic} dependence on $B_{\parallel}$, that
is, it can increase and then decrease as $B_\parallel$ increases.
We find that the enhancement (suppression) of $T_{KT}$ as $\Theta$
increases can be large for the system near the C-S (F-C) phase
boundary. However, when the system lies between the two phase
boundaries, the magnitude of $T_{KT}$ may have roughly the same
value. Since the crossed line is the predicted $T_{KT}$ for the
sample studied in Ref.[\onlinecite{Pellegrini2}], which has a
decreasing dependence on $\Theta$, the experimentally observed
enhancement of $T_{SDE}$ can not be explained by the result of
$T_{KT}$.

%\section{Hartree-Fock Theory}

Motivated by the above results, we look for another characteristic
temperature above which the spectral weight of the $\omega_0$ mode
indeed becomes significant. As mentioned before, for the systems
near the F-C phase boundary in the quantum phase diagram, a
non-vanishing spin polarization is possible when $T \gtrsim
T_{KT}$. Consequently, the mean-field Hamiltonian of the
self-consistent Hartree-Fock theory at finite temperatures, which
takes this fact into account, reduces to
\begin{eqnarray}
H^{HF}&=&-\frac{\Delta_{\rm SAS}+\delta_{\rm SAS}}{2}
\sum_{\tau,k,\sigma}\tau c^\dagger_{\tau,k,\sigma}c_{\tau,k,\sigma}
\nonumber\\ &&- \frac{\Delta_z+\delta_z}{2} \sum_{\tau,k,\sigma}
\sigma c^\dagger_{\tau,k,\sigma}c_{\tau,k,\sigma},
\end{eqnarray}
where $\delta_{\rm SAS}=(U_E/2) \sum_{\tau,\sigma} \tau
f(E_{\tau,\sigma})$ and $\delta_z=(U_A/2) \sum_{\tau,\sigma}
\sigma f(E_{\tau,\sigma})$. Here the Landau gauge is assumed, and
$c _{\tau,k,\sigma }^{\dagger}$ creates an electron at the lowest
Landau level orbital $k$ in the symmetric ($\tau=1$) or the
antisymmetric ($\tau=-1$) subbands with spin $\sigma / 2$
($\sigma=\pm1$). The thermal averages $\langle
c^\dagger_{\tau,\sigma}c_{\tau,\sigma}\rangle=f(E_{\tau,\sigma})$,
where $f(E)$ is the Fermi-Dirac distribution function and the
energy eigenvalues of this mean-field Hamiltonian are
\begin{equation}\label{eigen}
E_{\tau,\sigma}=-\frac{\tau}{2}(\Delta_{\rm SAS}+\delta_{\rm SAS})-
\frac{\sigma}{2}(\Delta_z+\delta_z).
\end{equation}
We assume that the translational symmetry is not broken, thus
these expectation values have no intra-Landau level dependence.
Since we consider only the case of $T>T_{KT}$, the order
parameters for the C phase are dropped. Moreover, because of the
symmetry in the energy levels, the chemical potential is fixed at
zero for all temperatures. We see that the four energy levels for
the non-interacting electrons are shifted by the self-consistent
mean fields, $\delta_{\rm SAS}$ and $\delta_z$, both of which have
temperature dependence.

\begin{figure*}[h]
\centerline{\epsfxsize=7cm \epsfysize=7cm \epsfbox{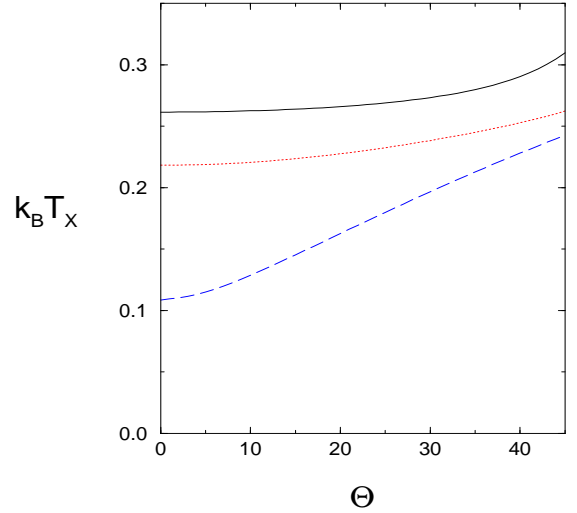} }
\caption{ $k_B T_X$ as a function of the tilted angle $\Theta$ (in
unit of degree) of the applied magnetic field with a fixed
$B_\perp$. The energy unit is $e^2/\epsilon l$. Continuous,
dotted, and dashed lines correspond to $\Delta_{\rm SAS}=$ 0.062,
0.15, and 0.23, respectively. The sample parameters, $\Delta_z,
d$, and $b$ are the same as in Fig.~\ref{fig:KT}. } \label{fig:Tx}
\end{figure*}

For $T >T_{KT}$, if we assume that the effective Zeeman energy,
$\Delta_z+\delta_z$, is initially larger than the effective
symmetric-antisymmetric energy gap, $\Delta_{\rm SAS}+\delta_{\rm
SAS}$, the crossing of energy levels can occur at a higher
temperature $T=T_X >T_{KT}$, because the mean field $\delta_z$ is
a monotonic decreasing function of $T$. Thus at $T=T_X$, where
level crossing occurs, one has
\begin{equation}\label{mfeq1}
\Delta_{\rm SAS}+\delta_{\rm SAS}=\Delta_z+\delta_z ,
\end{equation}
\begin{equation}\label{mfeq2}
{1\over2} \sum_{\tau,\sigma} \sigma f(E_{\tau,\sigma})
={1\over2}\sum_{\tau,\sigma} \tau f(E_{\tau,\sigma})
=f(E_{+1,+1})-{1\over2},
\end{equation}
By solving Eqs.~(\ref{mfeq1}) and (\ref{mfeq2}) with
Eq.~(\ref{eigen}), the level-crossing temperature $T_X$ can be
determined. Combining the knowledge of the KT transition
temperature,\cite{DasSarma1,DasSarma2} a complete phase diagram at
finite temperatures can be obtained as shown in
Fig.~\ref{fig:phase}, where the phase boundaries for the tilted
angle $\Theta=30^{\circ}$ are also presented. In the F phase, the
planar spins are thermally randomized but $\langle S_z \rangle$
remains nonzero. This finite-temperature phase diagram indeed
confirms our previous arguments. Note that for $\Delta_{\rm SAS}$
slightly larger than $0.23\,e^2/\epsilon l$, the C phase directly
transits to the S phase at finite temperatures, in which $\langle
S_z \rangle=0$. It can be seen that an in-plane magnetic field
moves the finite-temperature phase boundaries to the right, due to
the effective modification of the sample parameters given in
Eq.~(\ref{new}). With a fixed $\Delta_{\rm SAS}$, the change of
the transition temperatures for a tilted magnetic field with
$\Theta=30^{\circ}$ can be read out directly from
Fig.~\ref{fig:phase}. This finite-temperature phase diagram
indicates that $T_X$ is an increasing function of $\Theta$. By
using Eq.~(\ref{new}), the tilted-field dependence of $T_X$ is
explicitly shown in Fig.~\ref{fig:Tx}. The values of $\Delta_{\rm
SAS}$ are chosen such that the corresponding samples can undergo
both C$\rightarrow$F and F$\rightarrow$S phase transitions with
rising temperatures (see Fig.~\ref{fig:phase}), even though only
the F$\rightarrow$S transition temperatures $T_X$ are plotted. In
general, it takes a higher $T_X$ for a sample with a smaller
$\Delta_{\rm SAS}$ to transit to the S phase, which is reasonable
since the F phase becomes more stable for a larger ratio of
$\Delta_z/\Delta_{\rm SAS}$. The in-plane magnetic field always
elevates $T_X$, in contrast to its effect on $T_{KT}$, which is
more complicated as shown in Fig.~\ref{fig:KT}. The result shows
that for systems near the $T=0$ F-C phase boundary (say, for the
cross symbol in the inset of Fig.~\ref{fig:KT}), the
level-crossing temperature $T_X$ indeed increases with
$B_\parallel$ and should be identified as the experimentally
observed $T_{SDE}$.

%\section{concluding remarks}

Before closing this paper, some remarks are in order. First, we
would like to comment that the threshold temperature in the
experiment\cite{Pellegrini1} is $T_{SDE}\approx 0.5$ K, which is
considerably lower then the calculated $T_X\approx 17$ K using the
actual experimental sample parameters (see Fig.~\ref{fig:phase}
for $\Delta_{\rm SAS}=0.1\,e^2/\epsilon l$). Therefore, the
present theory is not quantitatively satisfying. The quantum
fluctuations neglected in the mean-field theory should lower the
calculated level-crossing temperature $T_X$ and reduce this
discrepancy. Although the above analysis is crude, it provides a
starting point for interpreting the enhancement of the threshold
temperature in Ref.[\onlinecite{Pellegrini2}]. Second, we notice
that, for the sample in the transport experiment\cite{Sawada2}
(say, the sample with the total density $n_t=0.7\times 10^{11}$
cm$^{-2}$ at the balanced point), which is initially located in
the C phase and near the $T=0$ F-C phase boundary, its activation
energy decreases as the tilted angle increases from zero. We
suggest that the energy scale set by $T_{KT}$ (i.e. the
vortex-antivortex binding energy) may be related to this
activation energy, since they have similar
$B_\parallel$-dependence.

In conclusion, we have investigated the dependence of phase
transition temperatures, $T_{KT}$ and $T_X$, on the in-plane
magnetic field and demonstrated that it is $T_X$, rather than
$T_{KT}$, that agrees qualitatively with the experimentally
reported threshold temperature. We have also obtained a
finite-temperature phase diagram of the bilayer systems based on
the Hartree-Fock approximation. A verification of these two
different phase transitions awaits experimental measurements to
probe the C phase more directly at lower temperatures. For
example, the heat capacity measurements should show power law
temperature dependence in the C phase (see Fig.~12 of
Ref.~[\onlinecite{DasSarma2}]) because of the existence of the
gapless Goldstone mode due to spontaneous symmetry breaking. Once
that being achieved, it would be quite interesting for future
experiments to confirm the predicted non-monotonic
$B_\parallel$-dependence of $T_{KT}$.

\acknowledgments
M.F.Y. acknowledges financial support by the
National Science Council of Taiwan under contract No. NSC
89-2112-M-029-001. M.C.C. is supported by the National Science
Council of Taiwan under contract No. NSC 89-2112-M-003-006.

\newpage

%%%%%%%%%%%%%%%%%%%%%%%%%%%%%%%%%%%%%%%%%%%%%%%%%%%%%%%%%%%%%%%%%%%%%%%%%%%%%%%%%%%%

%%%%%%%%%%%%%%%%%%%%%%%%%%%%%%%%%%%%%%%%%%%%%%%%%%%%%%%%%%%%%%%%%%%%%%%%%%%%%%%%%%%%

\widetext


\begin{references}
\bibitem{note}
In contrast, the $\omega_0$ mode has a maximum spectral weight in the
S phase, where there are as many spin-up (down) empty states as there
are spin-up (down) electrons at low temperatures.
\bibitem{YC2}
For example, starting from the effective boson
theory,\protect\cite{Demler,KYang1} we get the same conclusions.
\bibitem{note1}
As discussed in Refs.~[\protect\onlinecite{DasSarma1,DasSarma2}],
the finite-temperature renormalizations for $T_{KT}$ can be much
larger in the vicinity of the C-S phase boundary such that this
estimation can then no longer be used. Since we are interested
mainly in the systems near the F-C phase boundary, our discussions
may be still valid.
\bibitem{note2}
Here we assume that two quantum wells are arbitrarily narrow, and the
effect of the in-plane magnetic field on orbital degrees of freedom
is ignored.
\end{references}
\end{document}